\documentclass{egpubl}
 
\JournalSubmission    
\usepackage[T1]{fontenc}
\usepackage{dfadobe}  

\usepackage{cite}  
\BibtexOrBiblatex
\electronicVersion
\PrintedOrElectronic
\ifpdf \usepackage[pdftex]{graphicx} \pdfcompresslevel=9
\else \usepackage[dvips]{graphicx} \fi

\usepackage{egweblnk} 

\title[Retrieval based Transfer Function Design]%
      {A Transfer Function Design Using A Knowledge Database based on Deep Image and Primitive Intensity Profile Features Retrieval}

\author[Y. Jung, J. Kong, \& J. Kim]
{\parbox{\textwidth}{\centering Y. Jung$^{1,2}$,
        J. Kong$^{1}$,
        and J. Kim$^{1}$
        }
        \\
{\parbox{\textwidth}{\centering $^1$School of Computer Science, the University of Sydney, Australia\\
         $^2$School of Computing, Gachon University, Republic of Korea\\
       }
}
}

\begin{document}


\maketitle
\begin{abstract}
   Direct volume rendering (DVR) is a technique that emphasizes structures of interest (SOIs) within an image volume visually, while simultaneously depicting adjacent regional information, e.g., the spatial location of a structure concerning its neighbors. In DVR, transfer function (TF) plays a key role by enabling accurate identification of SOIs interactively as well as ensuring appropriate visibility of them. TF generation typically involves non-intuitive trial-and-error optimization of rendering parameters (opacity and color), which is time-consuming and inefficient. Attempts at mitigating this manual process have led to approaches that make use of a knowledge database consisting of pre-designed TFs by domain experts. In these approaches, a user navigates the knowledge database to find the most suitable pre-designed TF for their input volume to visualize the SOIs. Although these approaches potentially reduce the workload to generate the TFs, they, however, require manual TF navigation of the knowledge database, as well as the likely fine tuning of the selected TF to suit the input. In this work, we propose a TF design approach where we introduce a new content-based retrieval (CBR) to automatically navigate the knowledge database. Instead of pre-designed TFs, our knowledge database contains image volumes with SOI labels. Given an input image volume, our CBR approach retrieves relevant image volumes (with SOI labels) from the knowledge database; the retrieved labels are then used to generate and optimize TFs of the input. This approach does not need any manual TF navigation and fine tuning. For our CBR approach, we introduce a novel volumetric image feature which includes both a local primitive intensity profile along the SOIs and regional spatial semantics available from the co-planar images to the profile. For the regional spatial semantics, we adopt a convolutional neural network (CNN) to obtain high-level image feature representations. For the intensity profile, we extend the dynamic time warping (DTW) technique to address subtle alignment differences between similar profiles (SOIs). Finally, we propose a two-stage CBR scheme to enable the use of these two different feature representations in a complementary manner, thereby improving SOI retrieval performance. We demonstrate the capabilities of our approach with comparison to a conventional CBR approach in visualization, where an intensity profile matching algorithm is used, and also with potential use-cases in medical image volume visualization where DVR plays an indispensable role for different clinical usages. 
   (see https://www.acm.org/publications/class-2012)


\end{abstract}  
\section{Introduction}

Modern imaging modalities, such as computed tomography (CT) or magnetic resonance (MR), is volumetric in nature, e.g., the shape, size, and location of structures can be natively described in three-dimensional (3D) space. They are represented as a stack of two-dimensional (2D) image slices that collectively show volumetric information. In conventional interpretation approaches, users manually navigate through an entire image volume, slice-by-slice, and then mentally reconstruct the volumetric information derived from 2D image slices. In an attempt to complement the 2D visualization, 3D rendering, such as direct volume rendering (DVR), is used to provide volumetric visualization of the data. Figure 1 shows the advantages of a 3D DVR for a human CT image volume compared to its 2D counterpart cross-sectional views, where the DVR provides an overview of the entire data and the shape of the kidneys and its spatial relationship with the neighboring spine and rib cage. Such visualization demonstrates the complementary view that is offered by the 3D DVR to the 2D cross-sectional views.

\begin{figure}[h]
    \centering
    \includegraphics[width=\columnwidth]{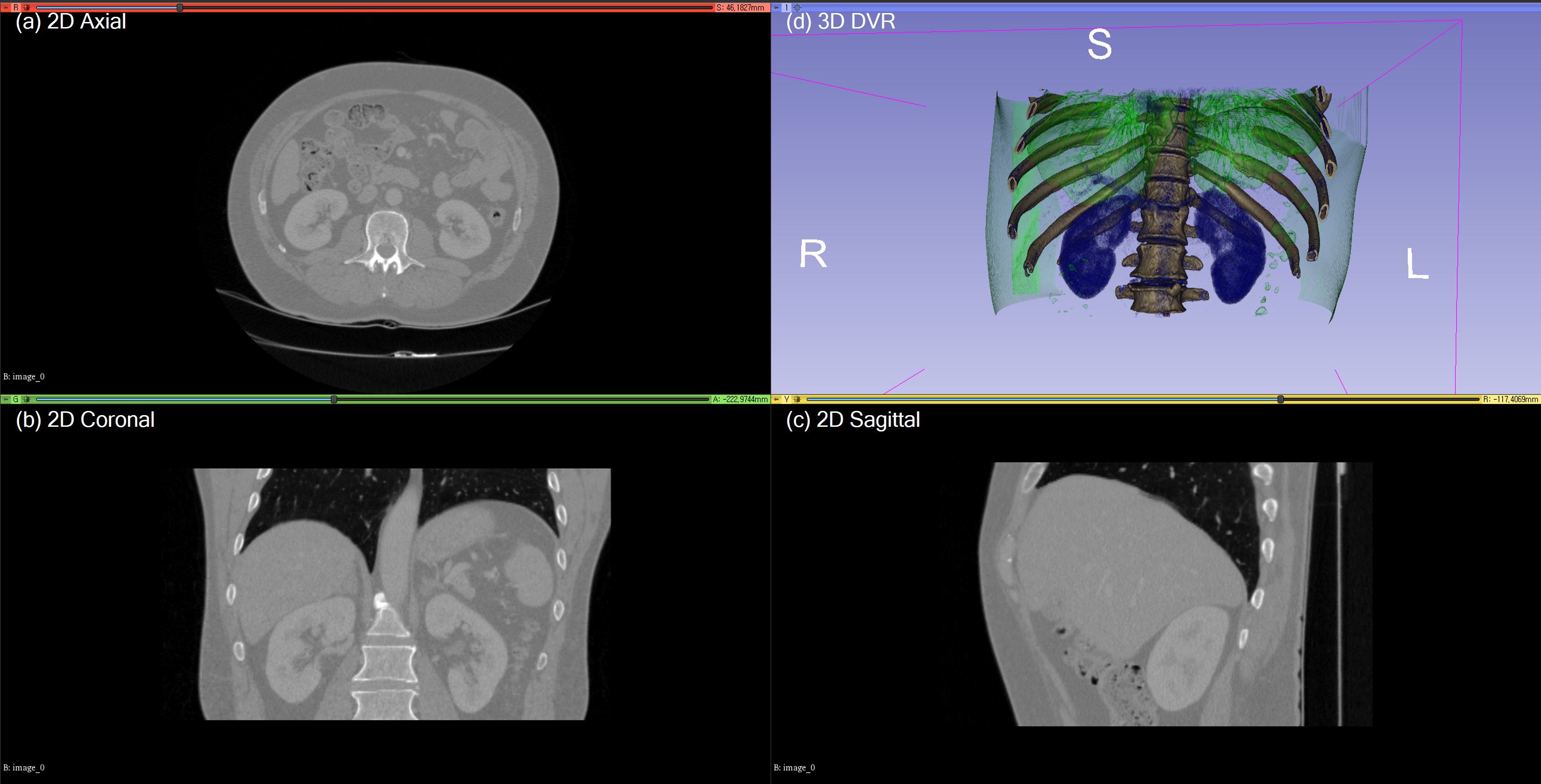}
    \caption{A quarter-view visualization example of complementary 2D and 3D for a human CT image volume. The three 2D cross-sectional views are shown in (a) axial, (b) coronal, and (c) sagittal views, respectively. 3D DVR of the same image volume is shown in (d). We used 3D Slicer software \cite{fedorov20123d} to generate the visualizations.}
    \label{figure 1}
\end{figure}

There are many challenges for DVR to be more broadly employed for volumetric imaging. A major technical issue lies with the data-specific visualization requirements, i.e., to identify and emphasize different structures of interest (SOIs) in an image volume. Transfer functions (TFs) play a key role in this \cite{ljung2016state}; for instance, in a conventional one-dimensional (1D) TFs that associate data density (intensity) with rendering parameters (color and opacity), a user must specify the intensity ranges (to identify SOIs) and then assign color and opacity values to the selected range (to emphasize the SOIs). As such, it involves non-intuitive trial-and-error adjustments of the TF parameters until the user ‘discovers’ the desired visualizations, which is time-consuming and inefficient \cite{pfister2001transfer}. There have been several works that have been directed at techniques to improve the TF design such as dominant approaches including multi-dimensional TFs \cite{kniss2001interactive, correa2009occlusion, correa2008size, caban2008texture}, image-centric \cite{ropinski2008stroke,guo2011wysiwyg, shen2014sketch}, and automated parameter optimization approaches \cite{he1996generation,correa2010visibility, jung2013visibility, jung2016efficient}. 

An alternative technique to TF design is with the use of ‘knowledge’ from similar cases \cite{marks1997design, guo2014transfer}, when compared to the other dominant approaches solely relying on input data. Pioneering works by Marks et al. \cite{marks1997design} and Guo et al. \cite{guo2014transfer} introduced a knowledge database that consists of a collection of pre-designed TFs by domain experts. In both works, instead of generating effective TFs from scratch, a user manually navigated the pre-designed TF cases in the database, with the visual aid of paired DVRs, and selected the most suitable one for the input image volume. Although these works demonstrated the feasibility of knowledge databases in TF design, they still required manual navigation of pre-designed TFs in the knowledge databases. In addition, pre-defined TFs from the knowledge database may not be always optimal for the input image volume. 

One solution to improve knowledge databases for TF design is to automate the manual searching process. Content based retrieval (CBR) is a technique that enables automated searching of similar cases against an input query by using content matching. The term 'content' refers to any features that can be derived from data itself for use in representing the data, e.g., with image data, conventional low-level features include intensity, texture, and color, with recent works reliant on deep learned features to better describe complex patterns and high-level semantics \cite{long2015fully, lecun2015deep}. CBRs have shown great success in general images, as well as with medical images, for their ability to effectively identify and retrieve similar cases from large databases \cite{kumar2013content, datta2008image, zin2018content}.

There has been a paucity of work on the use of CBR for TF design. As a pioneering and only relevant work, Kohlmann et al. \cite{kohlmann2009contextual} made use of an intensity profile, as content, to represent a SOI along a viewing ray. In their quarter-view visualization approach, a user-defined ray in 3D DVR was used to link a slice position of 2D cross-sectional views; the slice position was computed by automatically matching the intensity profile of the input ray query to pre-calculated profile templates for different SOIs in the knowledge database. As such, their use of CBR was highly optimized to provide complimentary 2D cross-sectional views for a single SOI, and is not suitable for TF design that associates with multiple SOIs. The proposed intensity profile matching only relied on low-level primitive features and may not be sufficient to identify and retrieve multiple SOIs. A conventional Euclidean Distance (ED) measure used, in addition, only considered voxel pairs that are exactly in the same position and could not match subtle differences (e.g., the two same SOIs but with different lengths).

In this work, we propose a new TF design approach using a knowledge database that is searched using a new CBR method, which we refer to as CBR-TF. Instead of pre-defined TFs \cite{kohlmann2009contextual}, our knowledge database consists of a collection of image volumes with SOI labels. Given an input image volume, our CBR-TF approach automatically retrieves similar cases by matching the content pair based on a volumetric image feature. The retrieved cases enable us to identify SOIs of the input, which is then used to generate a component-based 1D TF \cite{ropinski2008stroke}. The main technical contributions of our CBR-TF can be summarized as follows:

\begin{itemize}
  \item We propose a ‘triplet input query’ (TIQ) to formulate a volumetric image feature from an input image volume. A TIQ comprises (i) a user-defined ray of SOIs and, (ii) two co-planar images to the ray in 3D coordinates.
  \item For the two co-planar images, we propose the use of a pre-trained convolutional neural network (CNN) \cite{krizhevsky2017imagenet} to obtain image feature representations that carry high-level of regional spatial semantics. We further make use of an intensity profile along the SOIs for locating them, using dynamic time warping (DTW) technique \cite{keogh2005exact} to address subtle alignment differences between similar profiles.
  \item We propose a two-stage CBR method to enable the use of the two different types of CNN and DTW in a complementary manner through a rank-based sequential combination of the two individual retrievals.
  \item We adopt a ‘stroke-based’ image-centric approach \cite{ropinski2008stroke,guo2011wysiwyg, shen2014sketch} for simple and intuitive user TIQ interactions. Using a typical quarter-view visualization (as in Figure 1), in our user interface, a user only needs to draw a single ray (on 2D cross-sectional views) or select a single point (on 3D DVR) to derive the TIQ.
  \item For situations where particular SOIs are required to be emphasized, a visibility based TF parameter optimization \cite{correa2010visibility, jung2013visibility} can be applied to the TF design.
\end{itemize}
\begin{figure*}[t]
    \centering
    \includegraphics[width=0.7\textwidth]{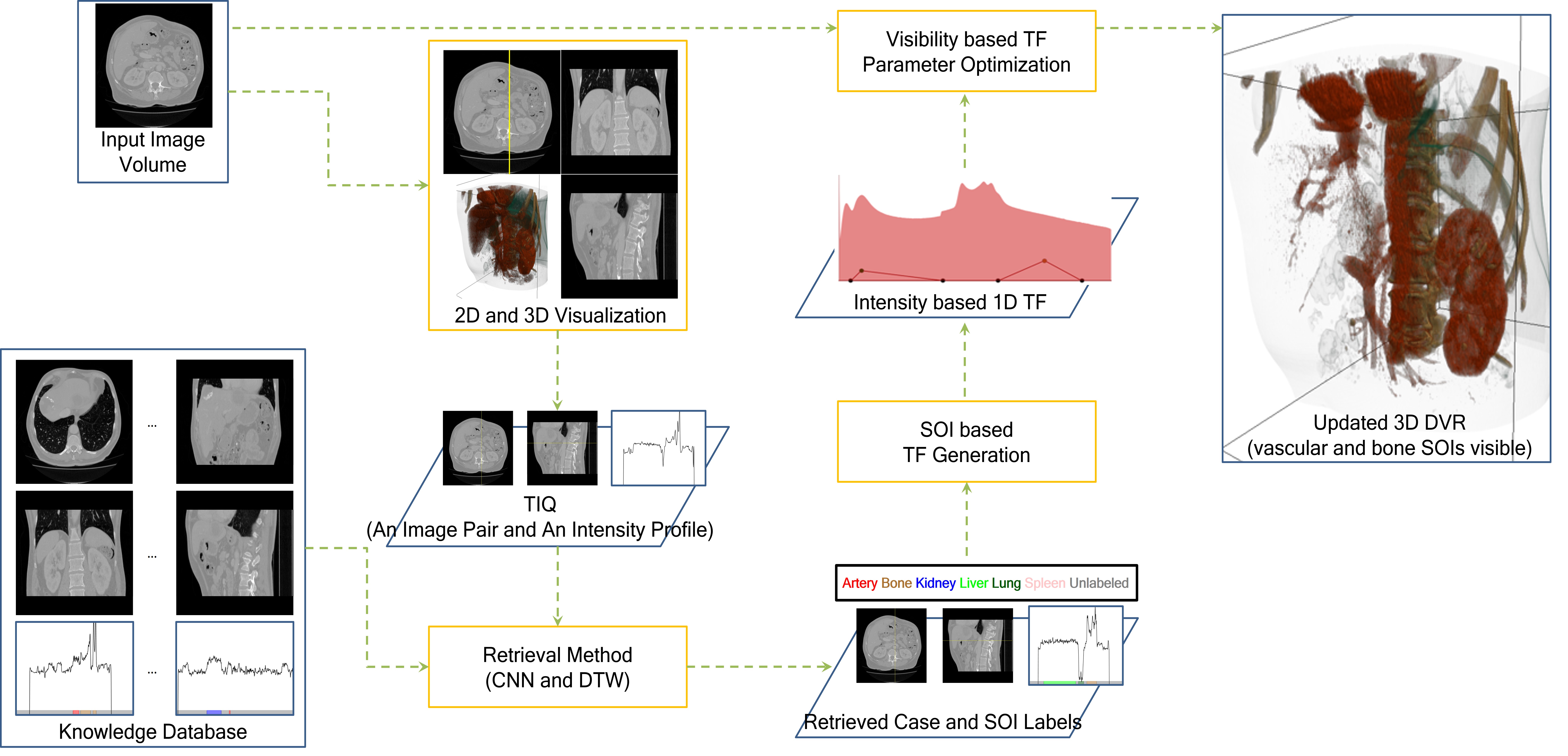}
    \caption{Overview of our CBR-TF approach.}
    \label{figure 2}
\end{figure*}
We demonstrate the application of our CBR-TF approach using a public medical imaging data repository of 3D Image Reconstruction for Comparison of Algorithm Database (3D-IRCADb-01) \cite{soler20103d}. This repository was selected for the availability of volumetric labels which are necessary to build our knowledge database of SOIs. We note that our CBR-TF approach can be applied to any knowledge database of image volumes that includes SOI labels; with our automated process to build the knowledge database, we can extend the current database when new labeled volumes are available.
\section{Related Works}
\subsection{2.1	Multi-dimensional TF Designs}
TF design has been advanced from traditional intensity based 1D TFs toward multi-dimensional TFs to facilitate the identification of SOIs. A pioneer work by Kindlmann et al. \cite{kniss2001interactive} proposed 2D TFs using intensity with its first or second order derivatives. Projecting these additional derived features as secondary dimensions on the TFs widget improved SOIs exploration; the new dimension typically acted as additional ‘indicators’ to guide users during TF design to help identify gradient information along the intensity (representing SOI) and visually emphasize the boundaries of SOIs. Similarly, Correa et al. \cite{correa2008size, correa2009occlusion} identified SOIs according to the local size \cite{correa2008size} or spatial relationship with its neighbors \cite{correa2009occlusion}. Some works used more than two dimensions (features), for example, the use of 20 local texture features by Caban et al.\cite{caban2008texture}, to enable greater differentiation among SOIs. As the number of TF dimensions increases, there is a greater need for manual optimizations in a complex multi-dimensional space \cite{pfister2001transfer, ljung2016state}. 
\subsection{Image-centric TF Designs}
Image-centric approaches make TFs design intuitive by allowing users to identify and optimize SOIs directly on an initial DVR visualization through manual gestures. Ropinski et al. \cite{ropinski2008stroke} enabled intuitive selection of SOIs by users drawing one or more strokes directly onto the DVR visualization near the silhouette of the SOIs. Then based on the stokes(s), an intensity histogram analysis was done to identify the intended SOIs in the intensity 1D TF widget. Guo et al. \cite{guo2011wysiwyg} manipulated the appearance of the intended SOIs through high-level ‘painting’ metaphors, e.g., eraser, contrast, and peeling. They used a 2D graph-cut algorithm to identify SOIs. Yuan et al. \cite{yuan2005volume} used 3D graph-cut algorithm for the SOI identification. Users, however, often has difficulty in precisely inferring SOIs from their interaction, especially, when the SOIs are fuzzy, semi-transparent, and multi-layered in the initial visualization. As such, these image-centric approaches may require manual and repetitive user interactions for desired visualizations. 
\subsection{Automated Parameter Optimization for TF Designs}
Some investigators \cite{he1996generation,correa2010visibility, jung2013visibility, jung2016efficient} focused on using parameter optimization algorithms to lessen the need for extensive user TF interaction. They assume if SOIs in a volume were known, the initial TF parameters could be automatically adjusted. These SOI based TF optimizations ensure that the visibility of the SOIs is maintained by reducing the opacity parameters of other structures / voxels that are occluding the SOIs. They generally relied on ‘greedy implementation’ that searches for locally optimal solutions, due to the unavailability of exhaustive search of huge TF parameters space. Such greedy solutions are sensitive to the initial parameters, which are often not properly defined by the users. A suboptimal initialization may result in the undesirable visualization of a local minimum.
\subsection{Knowledge Databases and CBR in DVR Visualization}
Marks et al. \cite{marks1997design} firstly generated a knowledge database with a collection of pre-designed TFs by domain experts for TF design. This database was comprised of pairs of a DVR and its corresponding TF. It was then used manually by the users to search through the TF cases that were most suitable to the input volume, with aid of paired DVRs. In another work, Guo et al. \cite{guo2014transfer} structured their knowledge database into a 2D search representation of TF cases via multi-dimensional reduction (MDS) for effective manual knowledge database navigation. In this search space, similar TF cases were grouped into clusters via the density field of their MDS. Although these works demonstrated the potential of how to use knowledge databases in TF design, they only partially reduced the workload to generate TFs as they required manual exhaustive navigations of the TF cases from the knowledge database, as well as the potential need for additional adjustment of the selected TFs to be best mapped to the input volume. 

Apart from TF design, Kohlmann et al. \cite{kohlmann2009contextual} investigated the use of a knowledge database to provide complementary visualization in 2D cross-sectional views and 3D DVR. They introduced CBR concept to automate search through the knowledge database, where the intensity profiles were used to represent the ‘content’, such as a SOI along the profile. This is the only work that demonstrated the use of the CBR based knowledge database for DVR visualization. 
\section{Methods}
\subsection{Overview}
An overview of our CBR-TF approach is shown in Figure 2 exemplified with an input image volume of the human upper-abdomen. We constructed, as an offline process, a knowledge database using a set of labeled image volumes (Section 3.2). In the quarter-view visualization of the input image volume with 2D cross-sectional views and a 3D DVR, a user drew a line (a ray) to visualize certain SOIs in any of the 2D views. With the user-selected ray, we generated a TIQ comprising a pair of co-planar images to the ray and its intensity profile. The TIQ was then used to query the knowledge database (Section 3.3). We used the SOI labels provided from the retrieved results to generate an intensity based 1D TF for the SOIs (Section 3.4), which can be further optimized with the input image volume to make sure that particular SOIs prioritize the visibility in the final 3D DVR (Section 3.5). 

\subsection{Knowledge Database Construction}

\begin{figure}[h]
    \centering
    \includegraphics[width=\columnwidth]{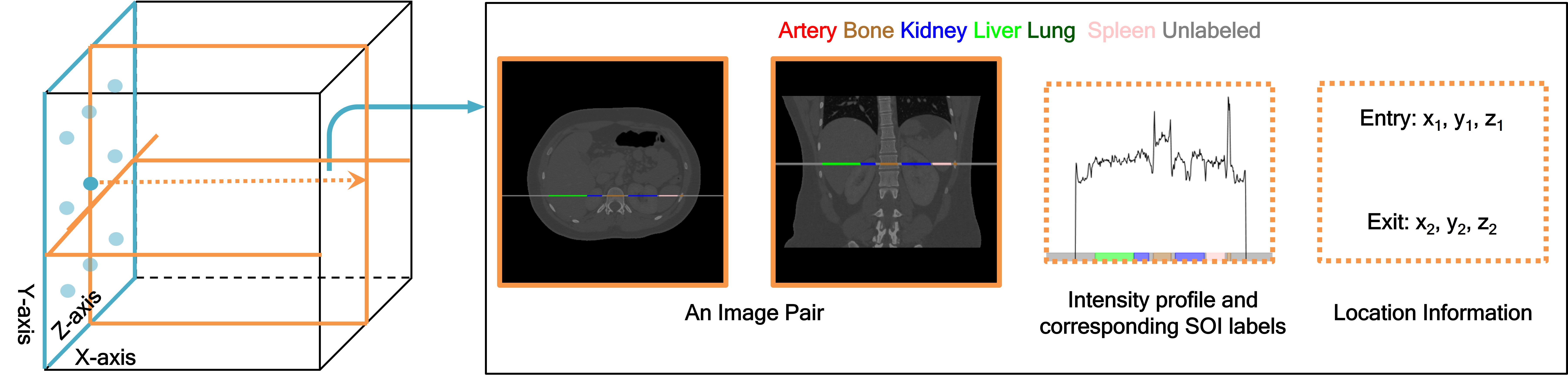}
    \caption{A ray extraction example and the ray representations (four items) used for our knowledge database construction. A point (cyan circle) is selected on a plane surface (cyan lines) of an image volume (black line cube) in the two primary axes (Y- and Z-axes). A ray (orange arrow dot line) is determined by casting it down to the third axis (X-axis). The four components were obtained as an image pair co-planar to the ray in X-Z axes and X-Y axes (two rectangles with orange lines), and an intensity and SOI label profile along the ray with its location coordinates.}
    \label{figure 3}
\end{figure}

We used a set of image volumes where every voxel had two information: intensity value and SOI label (name of structures in our setting) it represents. A primary element in our knowledge database is a ray. Ray extraction from an image volume is illustrated in Figure 3. We chose a point on a plane surface of the cube (image volume) aligned with two of the three primary axes (x, y, z) and cast a ray down to the third axis. We derived ray representations comprising four items from the extracted ray: (i) an image pair co-planar to the ray according to the primary axes; (ii) a profile of intensity values along the ray; (iii) the corresponding profile of the SOI labels; (iv) location coordinates of entry- and exit-point of the ray. We uniformly extracted multiple rays along a dimension (axis), with the consistent number of rays from all the three axes. Empty backgrounds of an image volume were excluded during the ray extraction. An image volume consisted of 192 rays in total and 64 rays from each of the three axes. The extraction interval, i.e., (the number of rays) was experimentally determined to reduce the likelihood of extracting highly similar rays as well as lowering the computational complexity of our CBR process. 

\subsection{Two-stage CBR Method}
For a TIQ from a user-selected ray q: a pair of images $q_i$ and an intensity profile $q_p$, we retrieved the most similar counterpart (c) from the knowledge database by calculating the similarity distances between the triplet pairs (q and c) in a two-stage manner. We first matched $q_i$ with $c_i$. Among the top $N$ retrieval results based on the image matching, as the second stage, we re-ranked them according to the similarity of intensity profiles ($q_p$ and $c_p$).

\textbf{CNN based Image Matching}. We used a well-established CNN, AlexNet, which was pre-trained using 1.2 million general images from the ImageNet database \cite{krizhevsky2017imagenet}. The architecture of AlexNet is shown in Figure 4. It consists of total 7 trainable layers with 60 million parameters; the first five are convolutional layers, and the remaining two are fully connected layers. The output of the last fully connected layer is a vector of 4096 dimensions. We used this as a feature vector for the image matching. 

\begin{figure}[h]
    \centering
    \includegraphics[width=\columnwidth]{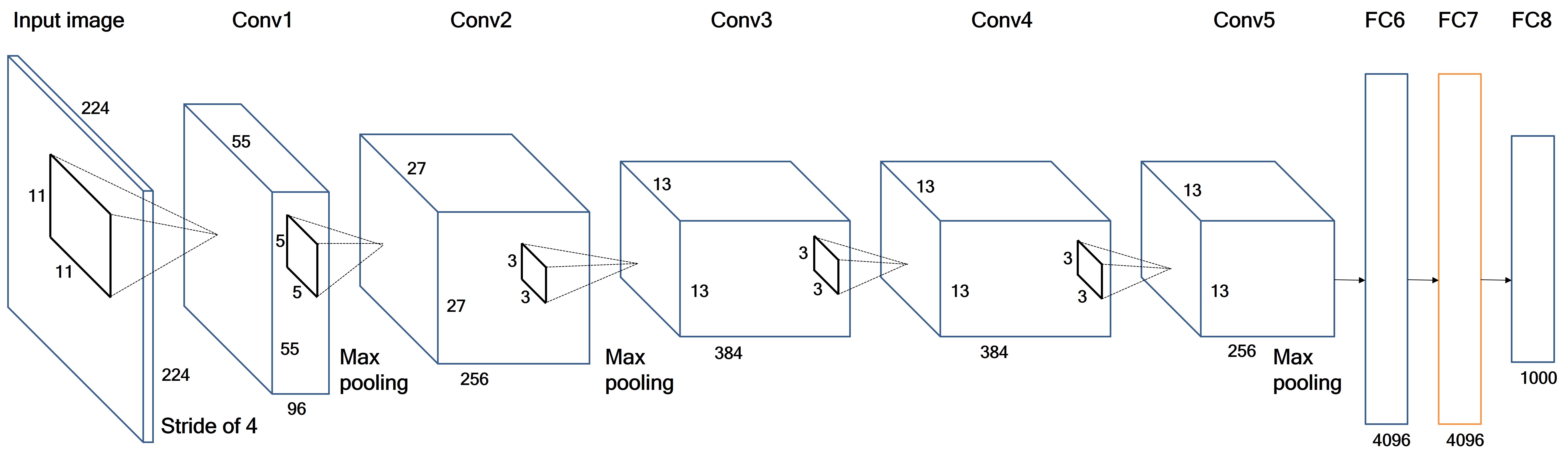}
    \caption{An illustration of AlexNet architecture. The first convolutional (Conv1) layer filters the 224 x 224 input image with 96 kernels of size 11 x 11 with a stride of 4 pixels. The second convolutional (Conv2) layer takes as input the (max pooled) output of Conv1 and filters it with 256 kernels of size 5 x 5. The third, fourth, and fifth convolutional layers are connected to one another without any pooling. The fifth layer is max pooled to the fully connected (FC) sixth layer. We used FC7 as a feature extractor for image matching.}
    \label{figure 4}
\end{figure}

We measured the image similarity distances by calculating Euclidean distances (EDs) between CNN image features extracted from $q_i$ and $c_i$. Each image from $q_i$ was only compared to the corresponding image from $c_i$ aligned with the same primary axes. We used the image similarity distance between $q_i$ and $c_i$: 
\begin{equation}\label{(1)}
    Image\:Similarity\:Distance (q_i,c_i) = \sum_{j=1}^{2}\sum_{b=1}^{4096}(F_{q_{i_j}}^b - F_{c_{i_j}}^b)
\end{equation}
where $q_{i_j}$ is $jth$ image from the input image pair, $c_{i_j}$ is $jth$ image from the image pair from the knowledge database, and $F^b$ is $bth$ CNN image feature.

\textbf{DTW based Intensity Profile Matching}. We measured the intensity profile similarity distance between $q_p$ and $c_p$ by calculating the cost of an optimal warping path using DTW. A $L_{q_p}\times L_{c_p}$ distance matrix was created which presets all pairwise distances between each element in $q_p$ and $c_p$, where $L_{q_p}$ is the length of $q_p$ and $L_{c_p}$ is to $c_p$. A warping path $P$ was then defined to pass through the contiguous low-cost parts of distance matrix to create a sequence of points =($p_1$,$p_2$,…,$p_K$) as shown in Figure 5. The optimal warping path $P^*$ was calculated as the path with the minimum cost:
\begin{equation}\label{(2)}
    Intensity\:Profile\:Similarity\:Distance (q_p,c_p) = min(\sqrt{\sum_{k=1}^{K}p_{k}}/K)
\end{equation}
where $K$ is the length of a sequence, and there are three constraints: (i) boundary condition: $P$ starts and ends in diagonally opposite corners of distance matrix; (ii) continuity: each step in $P$ proceeds from, and moves to, an adjacent cell; and (iii) monotonicity: $P$ does not take a step backwards in spatial locations.

\begin{figure}[h]
    \centering
    \includegraphics[width=0.4\columnwidth]{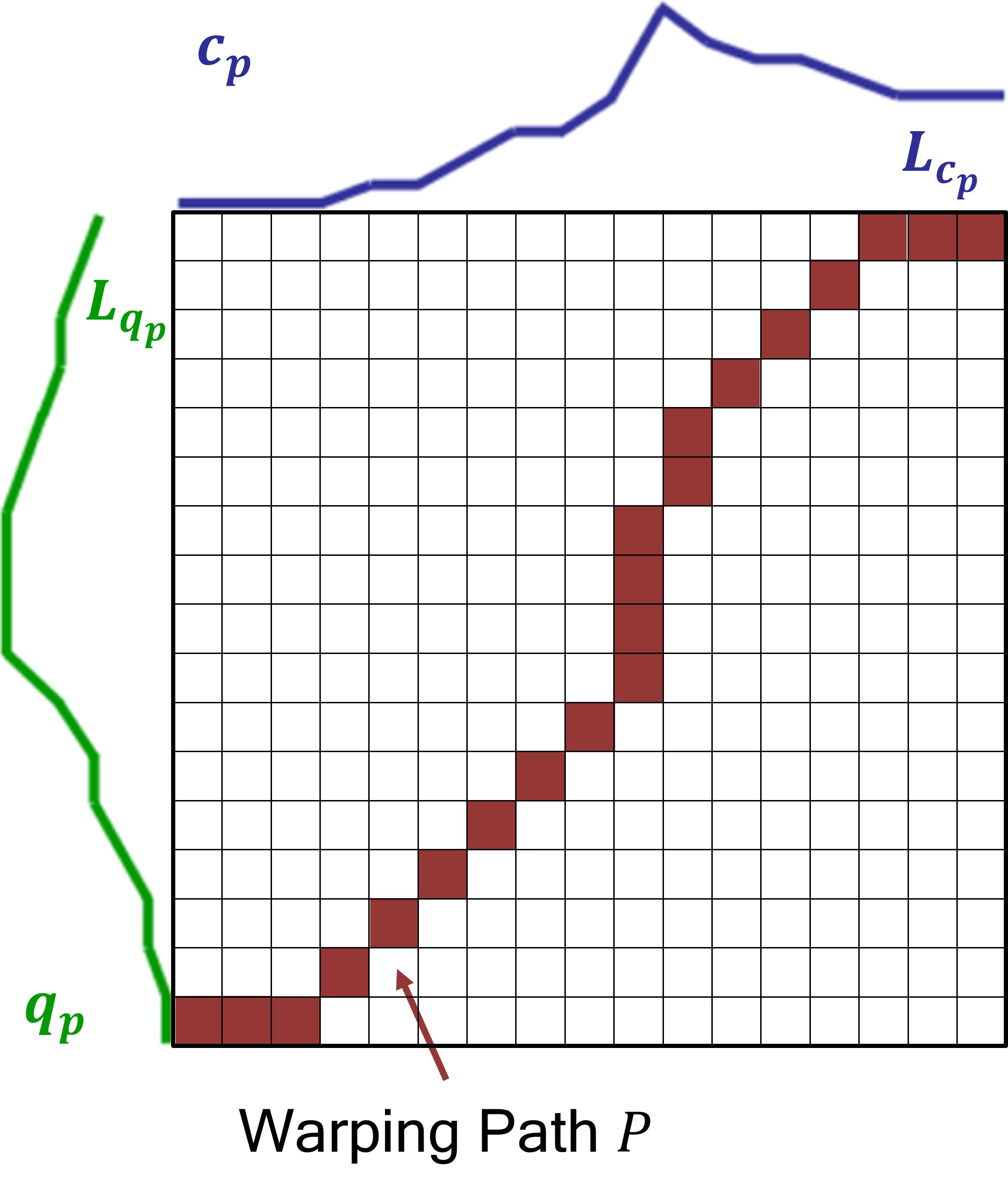}
    \caption{Warping path between an intensity profile pair example ($q_p$ and $c_p$) in DTW. }
    \label{figure 5}
\end{figure}

\subsection{SOI based TF Generation}
We generated a conventional intensity based 1D TF for SOIs of a TIQ using SOI labels retrieved from the knowledge database. We formulated a separate TF component $T_m$ for each SOI $S_m,m=1,…,M$ where $M$ was the number of SOI labels. Each component was a tent-shape in TF parameter space (intensity-opacity), and was specified as follows:
\begin{equation}\label{(3)}
    T_j=\{(I_{l,m},0),(\overline{I}_m,{\sigma_m}),(I_{h,m},0)\}
\end{equation}
where $I_{l,m}$ is the lowest intensity value in $S_m$, $I_{h,m}$ is the highest intensity value in $S_m$,  $\overline{I}_m$ is the average intensity in $S_m$, and $\sigma_m$ is the initial opacity of a tent peak set to 0.3 (where $\sigma_m$=1.0 indicates full opacity). The initial TF was given by the union of the tent shapes.
\begin{equation}\label{(4)}
    \cup_{m=1}^MT_m
\end{equation}
We used the tent shapes since they revealed the iso-surface of their corresponding SOIs, allowing a TF to visualize multiple SOIs in a semi-transparent manner. This is a desirable property in the visualization of complex SOIs. Nevertheless, our TF generation is not limited to this particular shape and can be re-parameterized to other shapes depending on specific visualization and data needs \cite{castro1998transfer}. We finally assigned each tent apex to unique colors using Color Brewer \cite{harrower2003colorbrewer} and all tent end points to a single color (black) to improve visual color depth perception among SOIs \cite{jung2018feature}. However, there is no limitation to apply different color schemes, e.g., the tent end points and apex can have the same color. 

\subsection{Visibility based TF Parameter Optimization}
Visibility based TF parameter optimization \cite{correa2010visibility, jung2013visibility} was an iterative process: the initial TF and target visibilities of SOIs were entered as the inputs. Each iteration calculated an intermediate new TF and the resulting visibility of SOIs. The visibility calculation is explained later in this section. The iteration continues until it converges to a minimum of an energy function E of visibility tolerance:
\begin{equation}\label{(5)}
    min_\theta\:E(\theta) = \sum_{m=1}^M(V_{T_m} - V_{R_m})^2
\end{equation}
where $M$ is the total number of SOIs, and $\theta$ is a set of opacity parameters of tent apexes forming the initial TF. The visibility tolerance is used to control how much the SOIs are visible in the optimized visualizations and it can be calculated via the sum of the squared differences between target visibility $V_T$ and resultant visibility $V_R$ for individual SOIs. The value of $V$ is the visibility proportion among all SOIs (where $V$ = 1.0 means the corresponding SOI exclusively priorities visibility); the assignment of $V_T$ was either automatically uniformly set to all SOIs or user-defined to enhance the visibility of particular SOIs.   

We used a downhill simplex method \cite{nelder1965simplex} to solve the optimization problem. This method is effective in a variety of practical non-linear optimization problems with multiple local minima \cite{barton1996nelder, lagarias1998convergence}. However, there is no limitation to apply other gradient-based approaches, e.g., gradient-descent \cite{ruder2016overview}. 

The visibility $F(p)$ for a voxel is the front-to-back opacity composition of all the voxels of a volume starting from a view-point $h$ to $p$ according to \cite{jung2013visibility}:
\begin{equation}\label{(6)}
    F(p) = e^{-\int_{p}^{h}A(g)dg}
\end{equation}
\begin{equation}\label{(7)}
    A(p)=A(p - \Delta p)+(1.0-A(p - \Delta p))O(p)
\end{equation}
where $A(p)$ is the composited opacity of the voxel coordinate $p$, $O(p)$ is its opacity, which is defined by a TF, and $\Delta p$ is the size of the sampling step. The visibilities of all voxels, weighted as the product of their opacity, are then added to determine the visibility of SOI (intensity range) given by:
\begin{equation}\label{(7)}
    V_{SOI} = \sum_{p\in SOI}O(p)F(p)
\end{equation}

\begin{table*}[h!]
  \begin{center}
    \caption{Recall metrics for retrieval among six different SOIs for the ED baseline \cite{kohlmann2009contextual} and our two-stage method with the two individual compartments.}
    \label{tab:table1}
    \begin{tabular}{|c|c|c|c|c|c|c|c|}
    \hline
     SOIs & Artery & Bone & Kidney & Liver & Lung & Spleen & All \\ 
    \hline
     ED (baseline) & 0.453 & 0.491 & 0.344 & 0.649 & 0.552 & 0.363 & 0.518 \\ 
    \hline 
     Our Two-stage & \textbf{0.612} & \textbf{0.692} & \textbf{0.648} & \textbf{0.778} & \textbf{0.829} & \textbf{0.644} & \textbf{0.719} \\  
    \hline
    \hline
     DTW (individual compartment) & 0.530 & 0.645 & 0.542 & 0.760 & 0.808 & 0.569 & 0.672 \\
    \hline 
     AlexNet (individual compartment) & 0.549 & 0.681 & 0.615 & 0.745 & 0.795 & 0.454 & 0.679 \\
    \hline      
    \end{tabular}
  \end{center}
\end{table*}

\begin{table*}[h!]
  \begin{center}
    \caption{Recall metrics for retrieval among six different SOIs for our two-stage method with three different types of pre-trained CNNs \cite{krizhevsky2017imagenet,szegedy2015going,he2016deep} in a comparison to the ED baseline \cite{kohlmann2009contextual}.}
    \label{tab:table2}
    \begin{tabular}{|c|c|c|c|c|c|c|c|}
    \hline
     SOIs & Artery & Bone & Kidney & Liver & Lung & Spleen & All \\ 
    \hline
     With Pre-trained AlexNet & \textbf{0.612} & \textbf{0.692} & \textbf{0.648} & \textbf{0.778} & 0.829 & \textbf{0.644} & \textbf{0.719} \\ 
    \hline 
     With Pre-trained GoogleNet & 0.591 & 0.681 & 0.616 & 0.770 & \textbf{0.836} & 0.608 & 0.707 \\  
    \hline
     With Pre-trained ResNet & 0.543 & 0.624 & 0.547 & 0.682 & 0.668 & 0.487 & 0.620 \\
    \hline 
     ED (baseline) & 0.453 & 0.491 & 0.344 & 0.649 & 0.552 & 0.363 & 0.518 \\
    \hline      
    \end{tabular}
  \end{center}
\end{table*}

\begin{table*}[h!]
  \begin{center}
    \caption{Recall metrics for retrieval among six different SOIs for the pre-trained AlexNet \cite{krizhevsky2017imagenet} and its fine-tuning counterpart \cite{shin2016deep}.}
    \label{tab:table3}
    \begin{tabular}{|c|c|c|c|c|c|c|c|}
    \hline
     SOIs & Artery & Bone & Kidney & Liver & Lung & Spleen & All \\ 
    \hline
     With Pre-trained AlexNet & \textbf{0.612} & 0.692 & \textbf{0.648} & \textbf{0.778} & \textbf{0.829} & 0.644 & \textbf{0.719} \\ 
    \hline 
      With Fine-tuned AlexNet & 0.609 & \textbf{0.700} & 0.617 & 0.765 & 0.816 & \textbf{0.688} & 0.714 \\  
    \hline
    \end{tabular}
  \end{center}
\end{table*}

\subsection{Evaluation Procedure}
3D-IRCADb-01 dataset \cite{soler20103d} was used for the evaluation. It consists of 20 CT image volumes of the upper-abdomen (10 males and 10 females). Since the dataset was mainly designed for benchmarking liver segmentation algorithms, there were limited manual labels available. We selected the six SOIs that commonly appear at the abdomen, which consist of the artery (1058 occurrences among all the images slices), bone (1824), kidney (506), liver (1645), lung (1101), and spleen (306). 

We believe that our chosen SOIs have sufficient complexity to measure the retrieval performances since the multiple SOIs, e.g., kidney, liver, and spleen, are similar in image features such as intensity and shape, and accurate retrieval among the SOIs is challenging. In total, our knowledge database was made of 3840 ray elements derived from the 20 image volumes, with each having 192 rays.

We compared our CBR-TF approach with the approach published in the visualization community by Kohlmann et al. \cite{kohlmann2009contextual}. This baseline approach relied on intensity profile matching using ED for measuring the similarity between the input and the cases from the knowledge database. For this comparison, only the second stage of our CBR-TF approach, which is the intensity profile matching using DTW, was applied in the knowledge database retrieval. We evaluated the first stage of our CBR-TF approach, which is image matching using pre-trained AlexNet \cite{krizhevsky2017imagenet}, by comparing to two established pre-trained CNNs of GoogleNet \cite{szegedy2015going} and ResNet \cite{he2016deep} by measuring the retrieval accuracy. The pre-trained AlexNet was also compared with the fine-tuned counterpart using a fine-tuning method \cite{shin2016deep}. We note that there are no comparable works that used image retrieval in the TF designs as well as DVR visualization.

The most common retrieval evaluation metrics of recall and precision were used. Recall refers to the number of times the SOI was correctly identified out of the number of times the SOI occurred. Precision refers to the number of times the SOI was correctly identified out of the number of times the SOI was identified by a retrieval approach. The values of both recall and precision range from 0.0 to 1.0 where a value of 1.0 indicates that all instances of a SOI type are identified by a retrieval approach. In our evaluation, recall was considered the most representative in evaluating retrieval performances; if false-positive occurs, the user can manually eliminate it in the subsequent TF design, but such manual optimization can be much more challenging in missing SOIs (i.e., false-negative). We, therefore, discussed the recall results and included the precision results in the Appendix (see Table A1). We conducted 10-fold cross validation for the evaluation, where for each fold, 18 CT image volumes were acted as the knowledge database, and the other 2 CT image volumes were used as the queries. We rotated this process 9 times to cover all 20 CT image volumes. 

For all these experiments, we used a consumer PC with an Intel i7 CPU and an Nvidia GTX 980 Ti GPU, running Windows 7 (x64), and implemented our CBR-TF approach using volume rendering engine (Voreen) \cite{meyer2009voreen}.

\section{Results}
\subsection{CBR Analysis}
We show the recall metrics for retrieval among six different SOIs between the ED baseline \cite{kohlmann2009contextual} and our two-stage method in Table 1. We also present results from using individual compartments of our method: DTW and AlexNet to analyze their individual performances and their contributions to the combined results. Compared to the ED baseline method \cite{kohlmann2009contextual}, our two-stage method outperformed in every SOI retrieval by 0.201 on average and kidney retrieval by 0.304 as the largest improvement. 

Our two-stage method outperformed both the individual compartments in all the SOIs. Within the two compartments, the AlexNet method was better than the DTW method for artery, bone, and kidney, and the other three SOIs were better retrieved by the DTW method. The DTW method improved upon the ED baseline counterpart across all the SOIs.

Our image retrieval compartment method is not limited to the particular CNN backbone of AlexNet \cite{krizhevsky2017imagenet} and can be applied to other CNNs. In Table 2, we showed the retrieval accuracies of pre-trained CNNs including AlexNet \cite{krizhevsky2017imagenet} and two other CNNs with deeper layers, GoogleNet \cite{szegedy2015going} and ResNet \cite{he2016deep}. Our results show that the pre-trained AlexNet, our default, had the highest average recall across all the SOIs except for lung where the pre-trained GoogleNet resulted in marginal improvement with 0.007. Regardless of the type of CNNs, they outperformed the ED baseline counterpart \cite{kohlmann2009contextual} across every SOI. The comparison result in Table 3 shows that the fine-tuned AlexNet was not always able to achieve higher recall accuracy compared to the pre-trained AlexNet; only spleen produced some meaningful improvement from the fine-tuned AlexNet by 0.044. This means that the fine-tuning method used \cite{shin2016deep} might not be optimal for our application.

\subsection{Qualitative Analysis of Our Two-stage CBR Method }

\begin{figure}[h]
    \centering
    \includegraphics[width=0.95\columnwidth]{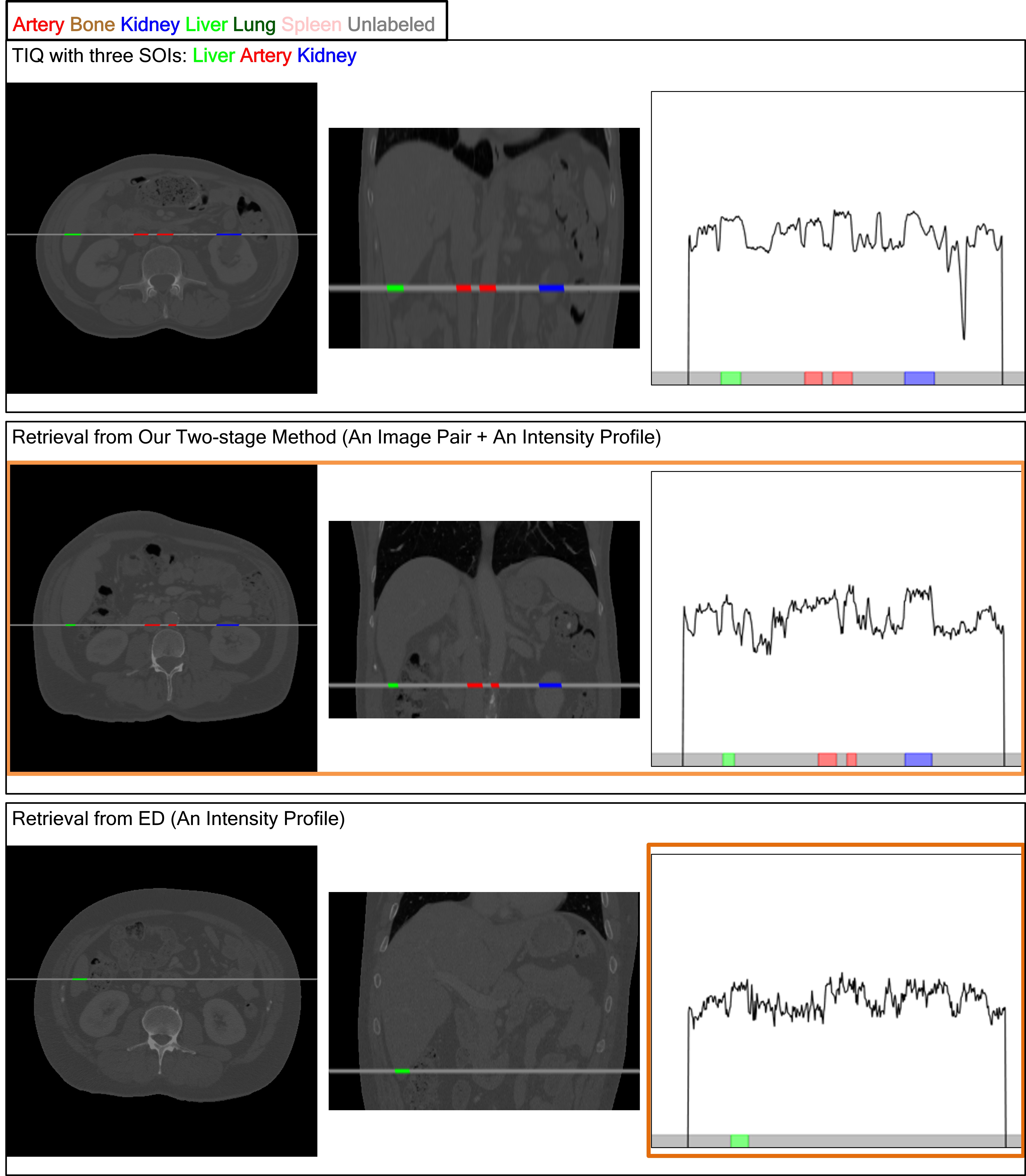}
    \caption{A comparison of our two-stage method (middle row) with the ED baseline method \cite{kohlmann2009contextual} (bottom row) using a TIQ (top row) that encounters three SOIs: the liver (bright green color), arteries (red color), and kidney (blue color). The right column shows the intensity profiles with the SOI labels, and the other two left columns show the associated image pair, respectively. The bounding boxes with orange lines indicate the compartments used for the retrieval computation.}
    \label{figure 6}
\end{figure}

Figure 6 exemplifies the capability of our two-stage method for SOI retrieval with comparison to the ED baseline method using only the intensity profile \cite{kohlmann2009contextual}. The results show that the ED baseline method incorrectly labeled the arteries and kidney along the TIQ. In contrast, all the SOIs along the TIQ were correctly retrieved when our two-stage method was applied.

\begin{figure}[h]
    \centering
    \includegraphics[width=\columnwidth]{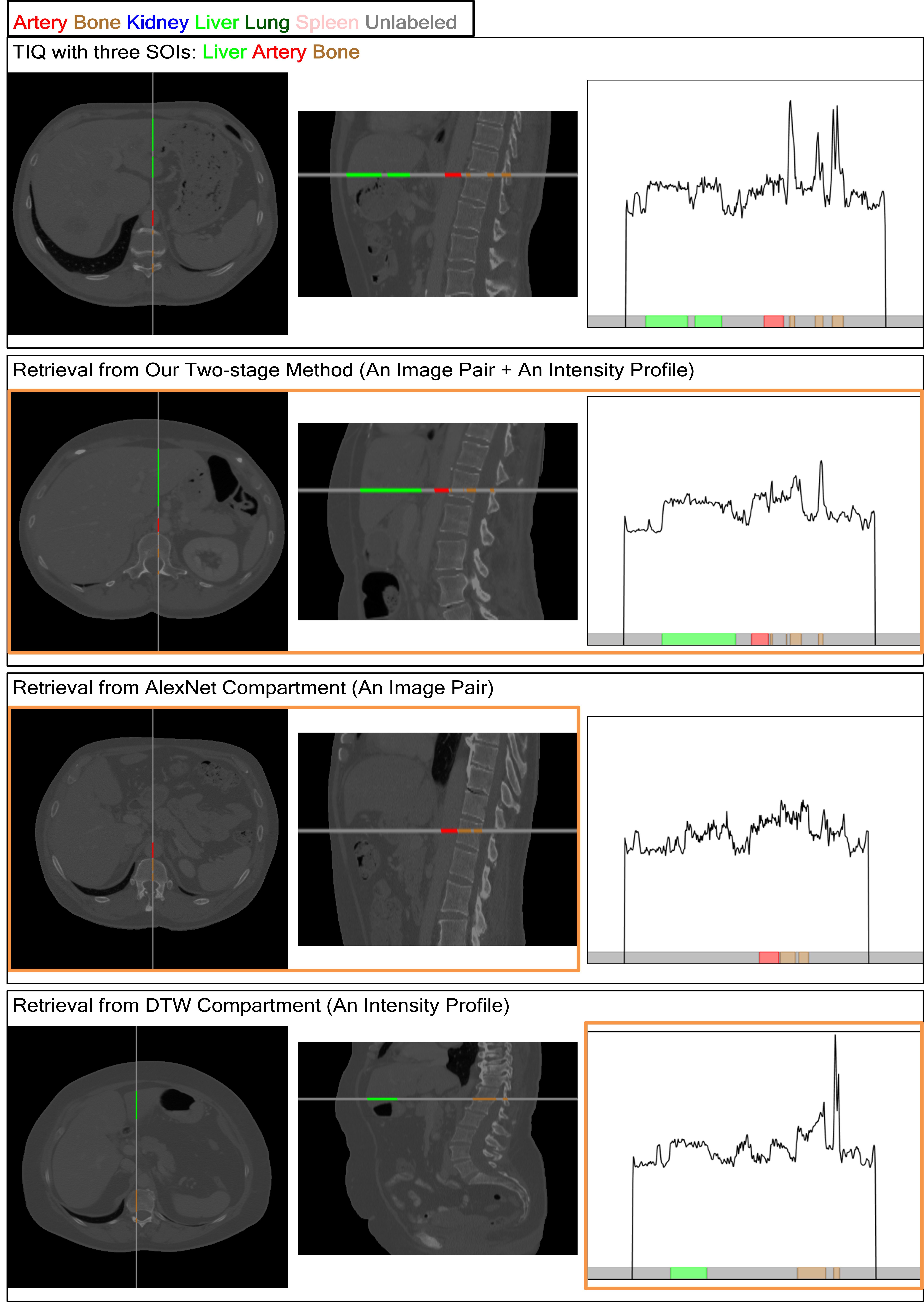}
    \caption{A comparison of our two-stage method (second row) with the two individual compartments: the image based AlexNet method (third row) and the intensity profile based DTW method (fourth row), with a TIQ (first row) that passes through three SOIs: the liver (bright green color), artery (red color), and bones (brown color). The bounding boxes with orange lines indicate the compartments used for the retrieval computation.}
    \label{figure 7}
\end{figure}

In Figure 7, we show the SOIs retrieval result from our two-stage method in a comparison to the two individual compartments. The results demonstrate that none of the compartment methods were sufficient for the adequate retrieval of SOIs present within the TIQ. In contrast, the two-stage method produced the same SOI labels as in the TIQ.

In the Appendix (see Figure A1), we included an intensity profile retrieval comparison between our DTW compartment and the ED baseline method \cite{kohlmann2009contextual} using another image volume data, where our DTW compartment method correctly identified all three SOIs while only one SOI was achieved in the ED baseline method. 

\begin{figure}[h]
    \centering
    \includegraphics[width=\columnwidth]{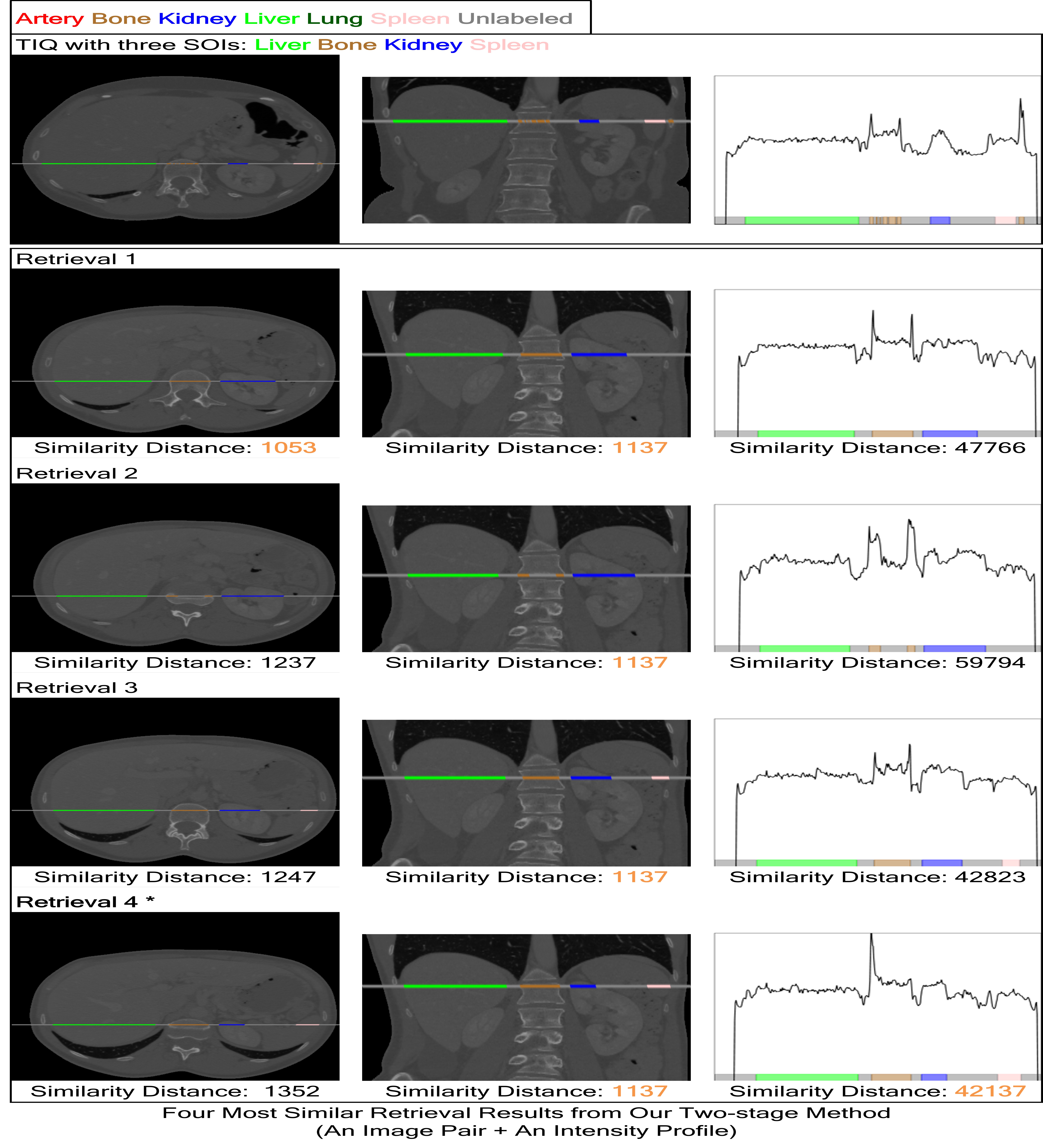}
    \caption{A complex TIQ (top row) that passes through four SOIs: the liver (bright green color), bone (brown color), kidney (blue color), and spleen (peach color) and the four most similar retrieval results from our two-stage method (other four rows, respectively). The smallest similarity distance (the best matching) is colored in orange for each compartment (each column). The last row is the overall best result from our two-stage method, denoted by ‘*’.}
    \label{figure 8}
\end{figure}

We illustrate the characteristics of our two-stage method, which searches across the entire image knowledge database, and then refines the retrieved results using the intensity profile matching. In Figure 8, we show the four most similar image retrieval results. We can see that these images were similar to the image query pair, with greater similarity (i.e., consistency in the same SOIs) in the coronal views (the second column). With the inclusion of the intensity profile matching results to the top-4 ranked image retrieval results, we see that the best intensity profile matching is with the $4th$ best image retrieval, among the only two results that were able to match the spleen (with the other result being the $3rd$ best image retrieval). Thus, in the combined result, the $4th$ image retrieval result will become the $1st$ ranked final retrieval result and has the overall best matching. 

\subsection{Automated TF Design Using Our CBR-TF Approach}

\begin{figure}[h]
    \centering
    \includegraphics[width=0.8\columnwidth]{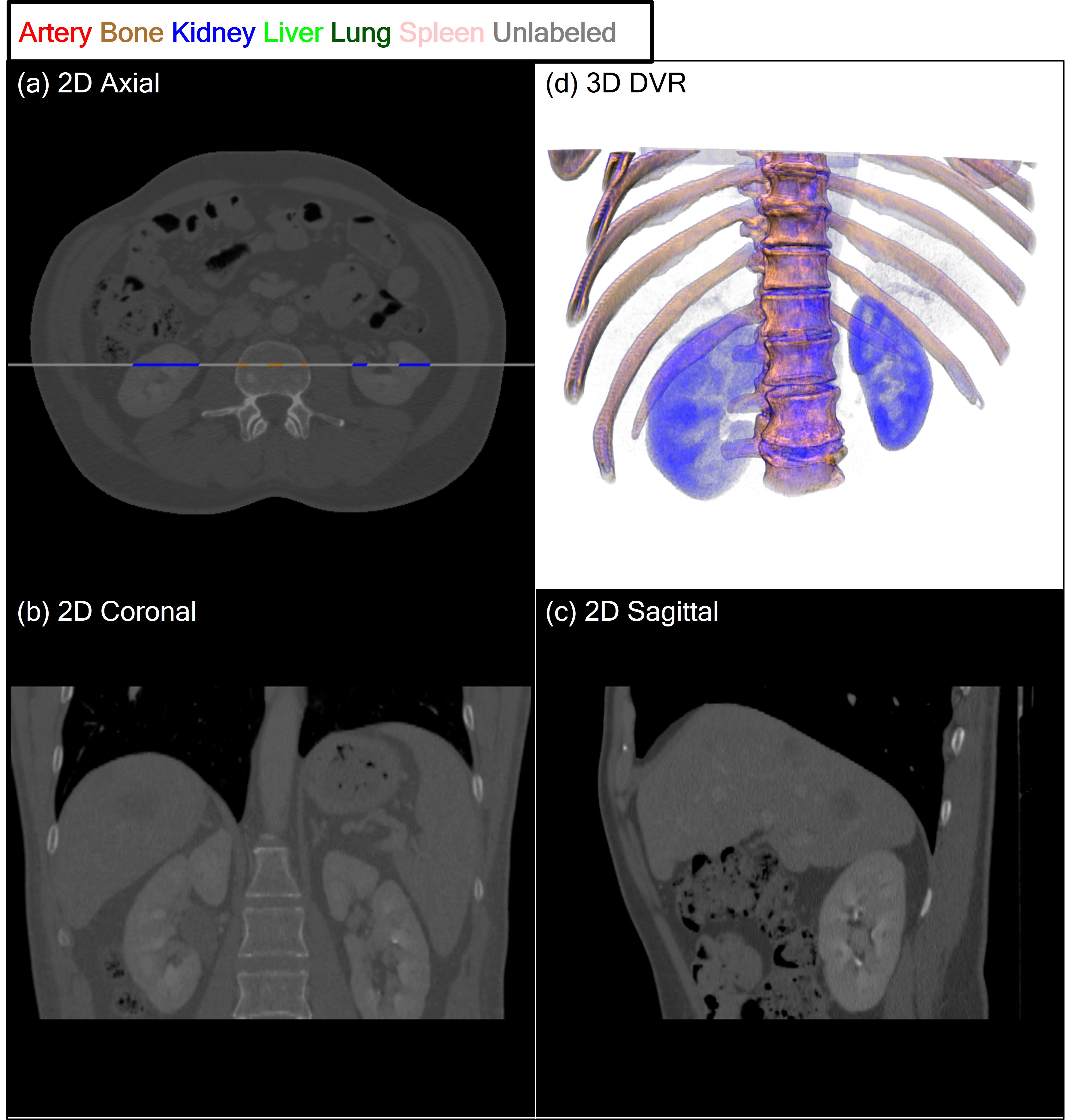}
    \caption{An example of a use-case for our CBR-TF approach, where the typical quarter-view visualization of a medical image volume is used: 2D cross-sectional (a) axial; (b) coronal; and (c) sagittal view, together with (d) a 3D DVR. A user specifies SOIs such as the kidneys (blue color) and the spine (brown color) by drawing a line containing them onto (a) the axial view (a ray with gray color) and our CBR-TF approach updates the DVR to emphasize them in 3D.}
    \label{figure 9}
\end{figure}

In Figure 9, a use-case of our CBR-TF approach is presented to effectively explore a medical image volume and generate a SOI-based DVR visualization. The user is only required to draw a line (a ray) onto any of the 2D cross-sectional views for our CBR-TF approach to identify and visualize SOIs. In this example, a line was drawn to contain the kidneys and bone in the axial view. An image pair and intensity profile associated with the ray were then used as the TIQ to retrieve annotations (labels) from the knowledge database to generate 3D DVR for the SOIs. The resultant DVR visualization provides the user with 3D information about the SOIs, including volumetric shapes and spatial relationships among the SOIs, which complements the 2D cross-sectional views. 

\begin{figure}[h]
    \centering
    \includegraphics[width=\columnwidth]{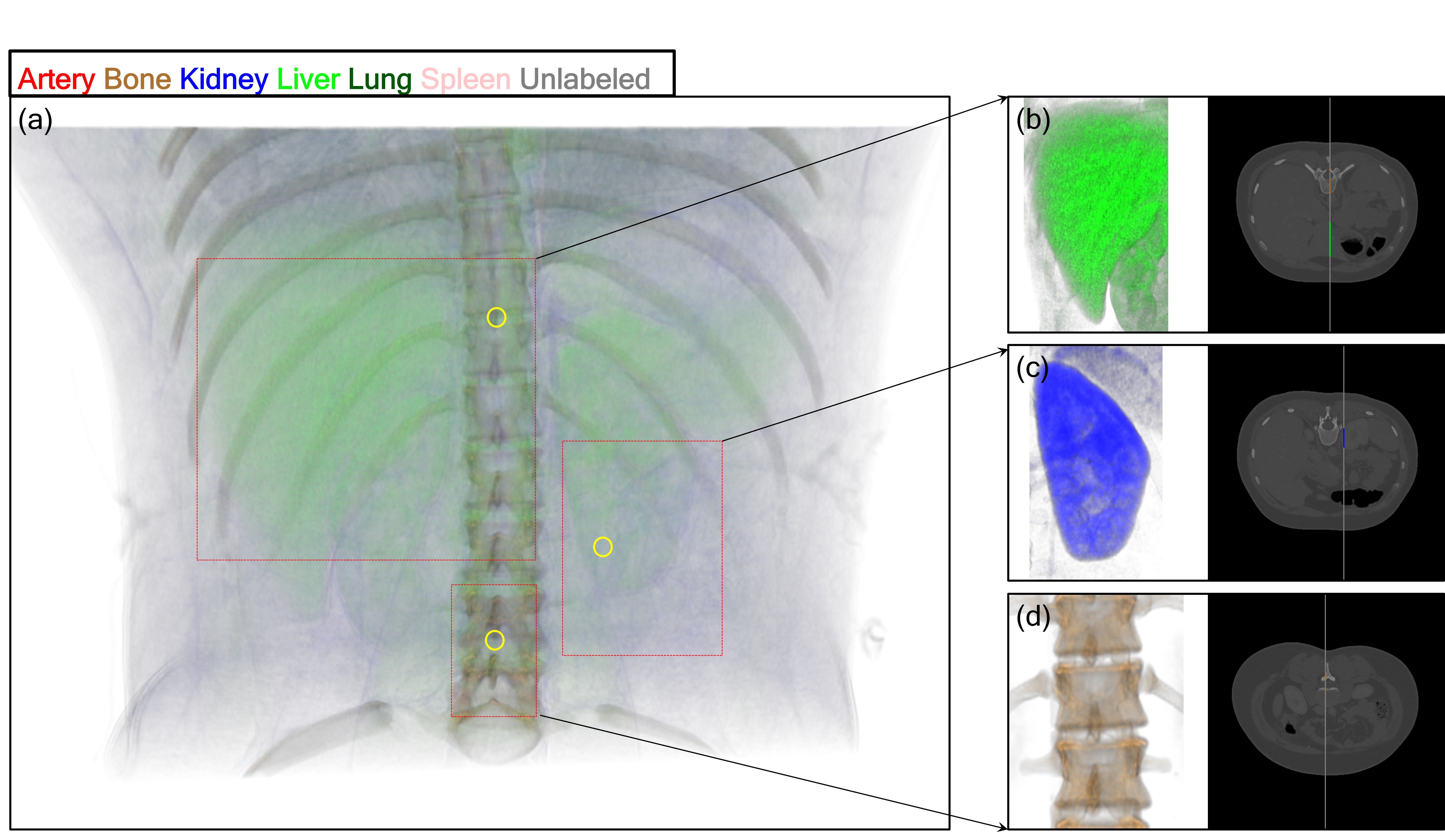}
    \caption{An example of a use-case for our CBR-TF approach for the visual enhancement of local SOIs directly from DVR visualization: (b) liver (bright green color); (b) kidney (blue color); and (c) bone (brown color), within (a) an upper-abdomen DVR. Each of the SOIs was selected by a single mouse click (yellow circles) on (a) the initial DVR where we constructed a viewing ray perpendicular to the camera origin which contains each SOI.}
    \label{figure 10}
\end{figure}

In Figure 10, we present another use-case of our CBR-TF approach. Instead of drawing a line in 2D cross-sectional views, the user can select a single point directly on an initial DVR visualization. From the selected point, a viewing ray perpendicular to a camera origin was constructed and used to generate a TIQ of an image pair and an intensity profile. This example illustrates how different SOIs can be identified based on the simple point selection, and how the TIQs can be used to automatically design the TF parameters such as to emphasize the local SOIs: (b) the liver, (c) the kidney, and (d) the bone. 

\begin{figure}[h]
    \centering
    \includegraphics[width=\columnwidth]{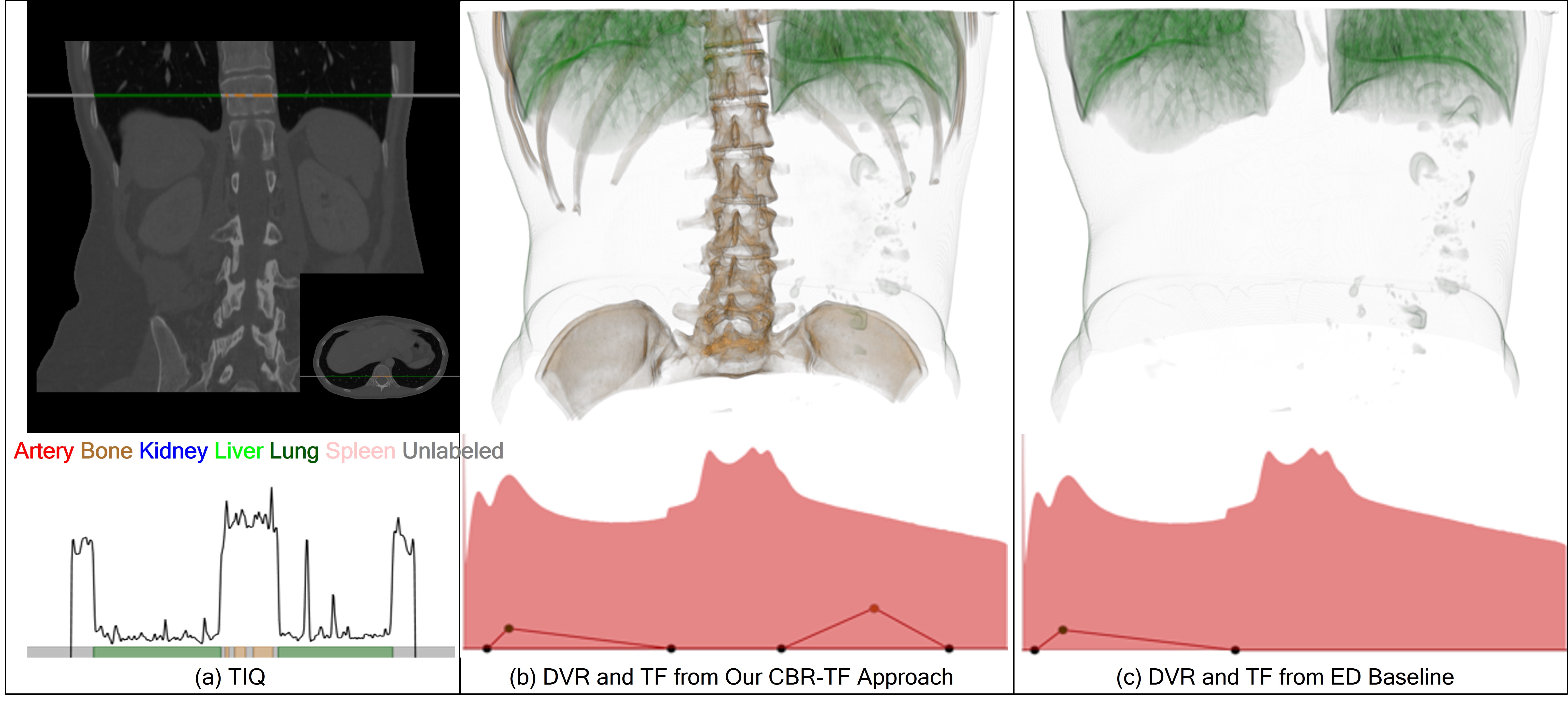}
    \caption{A comparison of TF and DVR visualization results from: (b) our CBR-TF approach; and (c) ED baseline approach \cite{kohlmann2009contextual} using a TIQ that consists of two SOIs of the lungs (dark green color) and bones (brown color) shown in (a).}
    \label{figure 11}
\end{figure}

We compared the TF generation results between our CBR-TF approach and the ED baseline counterpart \cite{kohlmann2009contextual}, in Figure 11. The results show that our CBR-TF approach produced the enhanced DVR outcomes, i.e., the correct identification of the SOIs, in comparison to the ED baseline counterpart. With the ED baseline , the bone failed to be identified, thus resulting in the missing TF component. 

\subsection{Computational Performance}

\begin{table}[h!]
  \begin{center}
    \caption{Average time (in seconds) to calculate the retrieval results from the knowledge database among the four different CBR methods.}
    \label{tab:table4}
    \begin{tabular}{|c|c|}
    \hline
     CBR methods & Time (seconds) \\ 
    \hline
     ED (baseline) & 00.84 \\ 
    \hline 
     Our Two-stage & 15.93 \\  
    \hline
    \hline
     DTW (individual compartment) & 01.14 \\
    \hline 
     AlexNet (individual compartment)& 15.70 \\
    \hline      
    \end{tabular}
  \end{center}
\end{table}

Computational time (in seconds) among the ED baseline \cite{kohlmann2009contextual} and our two-stage method for retrieval from the knowledge database is presented in Table 4. Our DTW compartment was marginally slower (30 milliseconds) than the ED baseline counterpart. As expected, the greater computation was required for the retrieval of images (AlexNet) compared to intensity profiles (DTW). In addition, TF generation computation time to obtain final DVR visualizations was < 1 second in all the Figures in this work.

\section{Discussions and Future Works}
The results demonstrated that our CBR-TF was able to automatically identify SOIs and generate TFs for an input image volume. We also showed simple and intuitive usability of our CBR-TF. 

\subsection{CBR}
We attributed our superior retrieval performance to our two-stage method, which utilized a rich volumetric image feature derived from our TIQ : (i) DTW characterized local primary intensity profiles from the ray of a TIQ, and (ii) regional high-level spatial semantics extracted by a CNN from the two co-planar images along the ray, in a complementary manner. The results demonstrated that none of the individual compartments could outperform our combined use  (see Figure 7). Our two-stage method was particularly effective in the cases where individual compartments produced the poor performance, such as artery, kidney, and spleen SOIs (see Table 1). This was consistent with the findings from existing works such as \cite{asadi2013effectiveness,ahmed2019content,wang2019enhancing}, which demonstrated the advantages in retrieval performance when leveraging two different yet complementary feature sets.

Our two-stage method complimented the intensity profile retrieval \cite{kohlmann2009contextual}, with co-planar image retrieval. One of the benefits of introducing image retrieval is that it could provide regional image semantics, e.g., not only shapes or textures of SOIs as well as the spatial relationship among the SOIs. This contributed towards differentiating the SOIs that have similar intensity values, and this differentiation was not able to be with the intensity profile retrieval only (see Figure 6 with the SOIs of the liver, artery, and kidney). 

Our results from DTW-based intensity profile retrieval outperformed the ED baseline counterpart \cite{kohlmann2009contextual}. The ED baseline only considered voxel pairs that were exactly in the same position and could not match subtle differences. Our DTW-based method addressed timing differences by warping the primary axis of one so that the maximum coincidence was attached with the other. This characteristic is important for our medical imaging application where the extents of the same SOIs (the intensity profiles) are widely varied among the patient image volumes. In such a scenario, a retrieval method must be able to accommodate these variations. A representative example is shown in the Appendix (see Figure A1).

For image retrieval, we used a pre-trained AlexNet \cite{krizhevsky2017imagenet} due to its highest average recall of 0.719 when compared to the two pre-trained deeper CNNs, GoogleNet \cite{szegedy2015going} and ResNet \cite{he2016deep} (see Table 2). This could be because the deeper networks learned more data-specific features that are relevant to general images, i.e., less generalizable to medical images. 

Our two-stage method, i.e., rank-based sequential combination of the two individual compartments, was experimentally designed based on the observation that the two individual compartments could be complimentary, but this advantage was not properly encoded when two individual compartments were used at the same time (i.e., a single-stage retrieval). In the first stage of image retrieval, we used the top 40 retrieval results (i.e., $N$=40), with the second stage of intensity profile retrieval only executed within the $N$ retrievals. This was empirically derived with variations in $N$ having a minor impact on the overall retrieval performances. Full details of this experiment were included in the Appendix (see Table A2). 

All the results presented in this work used a well-established recall metric to evaluate the retrieval performances. As a complementary metric, we also included precision results in the Appendix (see Table A1); we note that, as with recall, our two-stage method performed better in precision compared to the ED baseline \cite{kohlmann2009contextual}.

\subsection{TF Design}
Our results showed that our image-centric interaction for TF design is intuitive and effective. Our CBR-TF approach required a user to simply draw a line in 2D cross-sectional views (see Figure 9) or a point in 3D DVRs (see Figure 10) to select SOIs. The usability of such image-centric interaction to directly select SOIs in a visualization space, not in an indirect TF widget space, was investigated by the early work \cite{ropinski2008stroke}, where positive informal feedback from clinicians who conducted interaction using a medical CT image volume was observed. 

The use-case with the quarter-view visualization (see Figure 9) can be effective in medical image visualization where the user, i.e., clinicians, generally has prerequisite knowledge of their data; they can readily define a line (a ray) with SOIs on the 2D cross-sectional views. The fact that they are a novice user for TF design makes our CBR-TF approach more valuable. They do not need to rely on the TF widgets that are complicated to and unfamiliar with them. Instead, the SOIs can be simply defined in the 2D cross-sectional views where they are usually working on in the clinical routine. 

With another use-case (see Figure 10) where the single point selection was directly done in an initial 3D DVR for the SOI identification, we suggested that our CBR-TF approach enables a user to obtain a more detailed visualization of specific SOIs intuitively and automatically, thereby easily inspecting information that might be hidden in the initial DVR. 

In our CBR-TF approach, we adopted an automated visibility based TF parameter optimization to ensure that some of the identified SOIs prioritize visibility in resulting DVRs. The user, therefore, can avoid additional manual TF fine tuning that may be required for the cases where the TFs from knowledge databases are not optimal for the particular SOIs of the input image volume. A TF parameter optimization application example is shown in Figure 11, where the visibility of the bones was prioritized over the lungs by setting $V_T$ of the bones to 0.7 and $V_T$ of the lungs to the rest (0.3). 

In our CBR-TF approach, TFs were comprised of individual tent-shape components to represent each SOI. In the cases when the false-positive from our SOI retrieval occurs, we note that the user needs to manually eliminate the TF component of the false-positive SOI. Such manual TF optimization can be more challenging in missing structures (i.e., false-negative). However, our CBR-TF approach is still effective because the user does not need to start the TF design from scratch, and it is easier to add new SOIs to the TF that already contains other SOIs. 

In terms of the generation of knowledge databases, existing approaches \cite{kohlmann2009contextual,guo2014transfer} collect pre-designed TFs by domain experts, whereas our CBR-TF approach relies on SOI labels of raw image volumes. In terms of the perspective of experts in knowledge database generation, generating TFs may take much effort compared to SOI labels. 

\subsection{Future Works}
The image retrieval results (Table 2) demonstrated the adaptability of our CBR-TF approach to different CNNs. As such, our CBR-TF approach is not constrained to a particular CNN backbone. We note that recent advances of CNNs, e.g., for segmentation or classification tasks, demonstrated the capabilities to better describe image features to represent SOIs \cite{litjens2017survey,shen2017deep,chen2022recent,guo2016deep}. These backbones could be leveraged for our CBR-TF approach, but optimal adoption should be investigated to improve SOI retreival perforamance (see Table 3). We consider this challenging but interesting investigation as our future work.

Efficient retrieval computation is an important requirement for the practical application of our CBR-TF approach. In this work, we used an exhaustive search where every case was matched with a TIQ. This serial matching was sufficient for 3896 cases in our knowledge database. We believe that less than 20 seconds to generate TFs (and retrievals) from scratch (see Table 4) is an acceptable user experience. However, for larger knowledge database sizes, we will investigate adopting GPU based parallelism libraries \cite{heidari2013parallel,margara2011high} to speed up retrieval computational efficiency by enabling individual matching in parallel. Another approach is to construct a knowledge database hierarchically, e.g., using a set cover problem \cite{alon2006algorithmic}, for rapid but approximated retrieval.

Our CBR-TF approach does not require any dataset-specific parameter settings for the knowledge database construction. Furthermore, we automated the entire construction procedure including ray extraction, representation, and storage. Different types of imaging modalities can be easily integrated for different application domains. We limited the scope of this work to medical imaging data, but we suggest that our CBR-TF approach can be applied to non-medical imaging datasets. 

We demonstrated our CBR-TF approach with intensity based 1D TF design because they are commonly used in a wide range of visualization applications and have fewer variables that influence visualization experiments. Our CBR-TF approach, however, is not restricted to 1D TFs and could be adapted to multi-dimensional TFs \cite{kniss2001interactive, correa2009occlusion, correa2008size} for greater differentiation of SOIs. Huang et al. \cite{huang2003rgvis} proposed an approach to generate 2D TFs when the identification of SOIs in the 2D cross-sectional views are available. We regard this as an important new avenue of research.

\section{Conclusion}
We presented a new TF design approach by proposing an automated two-stage CBR for a knowledge database, and the use of its retrieved results in generating TFs. Our experimental results demonstrate that our CBR-TF approach was capable of identifying and visualizing SOIs from the combination of the complementary image semantics with the intensity profile matching. 

\bibliographystyle{eg-alpha-doi} 
\bibliography{EGauthorGuidelines-body}       


\end{document}